\documentclass[amsmath, amsfonts, showpacs, twocolumn, prl]{revtex4}
\usepackage{graphicx}
\usepackage{epsfig}
\usepackage{bm}
\usepackage{dcolumn}
\usepackage{amsmath}
\usepackage{amssymb}
\usepackage{euscript}

\begin{document}

\title{Probing the pseudogap phase of cuprates by a Giaever transformer}

\author{Alex Levchenko}
\affiliation{Materials Science Division, Argonne National
Laboratory, Argonne, Illinois 60439, USA}

\author{M. R. Norman}
\affiliation{Materials Science Division, Argonne National
Laboratory, Argonne, Illinois 60439, USA}

\begin{abstract}
We develop a theory of the rectification effect in a double-layer
system where both layers are superconductors, or one of the layers
is a normal metal. The Coulomb interaction is assumed to provide the
dominant coupling between the layers. We find that superconducting
fluctuations strongly enhance the drag conductivity, with
rectification most pronounced when both layers are superconductors.
In view of their distinct dependence on temperature near $T_c$ and
layer separation, drag measurements based on a Giaever transformer
could distinguish whether rectification occurs due to fluctuating
pairs or inductively coupled fluctuating vortices.
\end{abstract}

\date{January 31, 2011}

\pacs{74.40.-n, 74.72.Kf, 73.40.Ei, 74.78.Fk}

\maketitle

\textit{Introduction}-- The origin of the pseudogap phase in high
temperature cuprate superconductors is the subject of on-going
debate~\cite{Timusk}. One possible explanation of this
phase is the presence of preformed Cooper pairs
which deplete the density of states near the Fermi level~\cite{Varlamov}.
This is consistent with photoemission studies which reveal that the Fermi
surface breaks up into disconnected
arcs~\cite{Norman-Arcs} that should have profound implications for the
nature of transport in the pseudogap phase~\cite{LMNP}.

In this connection, recent observations~\cite{Xu,Wang}
of a large Nernst signal in the pseudogap phase has generated much
interest. Two competing theories may account for this
phenomenon. On the one hand, in proximity to the superconducting
transition temperature $T_c$, it is natural to attribute a large Nernst
response to fluctuating vortex-like excitations which carry entropy.
On the other hand, there exists a close
correspondence of the Nernst signal with the fluctuational
diagmagnetism~\cite{Li10}, which points towards the Nernst effect as
originating from fluctuating pairs. Existing theories based
on Ginzburg-Landau or diagrammatic
approaches~\cite{Uss02,Serbyn,LNV} give a good description of the
Nernst data. It is thus
desirable to have an additional probe of the pseudogap phase which can
distinguish transport contributions arising from fluctuating vortices
or preformed pairs. In this paper, we suggest using a Giaever
transformer~\cite{Giaever} for such a probe and develop the theory for this.

Measurements on double-layer
systems~\cite{Drag-2DEG-Exp} have provided us invaluable information
about the scattering mechanisms underlying transport. A
typical experiment consists of driving one of the subsystems
(active layer) and measuring an induced response in the other one
(passive layer). Physically, by driving a current in the active layer,
one creates an electromagnetic environment to which the passive
layer is exposed. These fluctuations are rectified, thus producing a
finite drag response. Under the condition of no tunneling between
the layers, momentum can be transferred only via electron-electron
scattering due to the inter-layer Coulomb interaction. In contrast,
if vortices are present, then
inductive coupling between the layers becomes important.
Experimentally, the measured quantity is the drag resistance, $R_D=V_D/I$,
which is the ratio between the induced voltage $V_D$ in the passive
layer to the driving current $I$ in the active one. The
temperature dependence of $R_D(T)$ is extremely sensitive to the
microscopic properties of the system. For example, in Fermi liquids
$R_D\propto T^2$, which results from the phase space
available for scattering. In contrast, in
disordered systems $R_D\propto T^2\ln T$~\cite{Drag-2DEG-Th}, while
for non-Fermi liquids, like composite fermions in quantum Hall
bi-layers, the drag exhibits a very different temperature
dependence, namely $R_D\propto T^{4/3}$~\cite{Drag-QH}. Many other
interesting examples can be found in the literature~\cite{Rojo}.

In recent work~\cite{Refael}, the rectification effect was studied in
the context of a magnetic field driven superconductor-insulator
transition and attributed to fluctuating vortices. With some
modifications, this theory can be applied to the case of cuprates
under consideration here. However, we take the point of view that the
dominant contribution to the rectification signal in the pseudogap
phase is due to fluctuating pairs. Below, we develop the corresponding
theory within a linear response Kubo formalism.

\textit{Formalism}-- We consider both a symmetric superconductor
bi-layer and a non-symmetric one where one layer is replaced by a
normal metal. The building blocks of the theory are the normal and
superconductor Greens functions
\begin{equation}\label{G}
G_n(p,i\varepsilon)=\frac{1}{i\bar{\varepsilon}_n-\xi_p}\,,\quad\!\!
G_s(p,i\varepsilon)=-\frac{i\bar{\varepsilon}_s+\xi_{p}}{\bar{\varepsilon}_s^{2}+\xi
_{p}^{2}+\Delta _{p}^{2}}.
\end{equation}
In what follows we assume a free electron dispersion relation,
$\xi_p=p^2/2m^*-\varepsilon_F$, which in two dimensions implies a constant density of
states, simplifying the energy integrations. We assume a momentum dependent
d-wave pseudogap of the form
$\Delta_{p}=\frac{\Delta}{2}[\cos(p_x)-\cos(p_y)]\simeq\Delta\cos2\vartheta_p$.
In Eq.~\eqref{G} $\bar{\varepsilon}_{n(s)}=\varepsilon
+\gamma_{n(s)} \,\mathrm{sign}(\varepsilon )$, with $\gamma_s\simeq
T$ being the scattering rate observed by photoemission in cuprates,
and $\gamma_n=1/2\tau_{n}$ a constant scattering rate assumed for the normal metal.

For the fluctuating Cooper pairs in a superconductor, we take the
standard expression for the propagator whose retarded component has
the form~\cite{book}
\begin{equation}\label{L}
L^R(Q,\Omega)=-\frac{1}{N_s}\frac{1}{\pi D_s
Q^2/8T+\epsilon-i\pi\Omega/8T+\Upsilon_\Omega}
\end{equation}
where $\Upsilon_\Omega=\frac{\Omega}{2\Delta}\frac{\partial
\Delta}{\partial\varepsilon_F}$ accounts for electron-hole
asymmetry, $N_s$ is the density of states, $D_s$ the diffusion
constant and $\epsilon=\ln(T/T_c)$.

\begin{figure}
  \includegraphics[width=8cm]{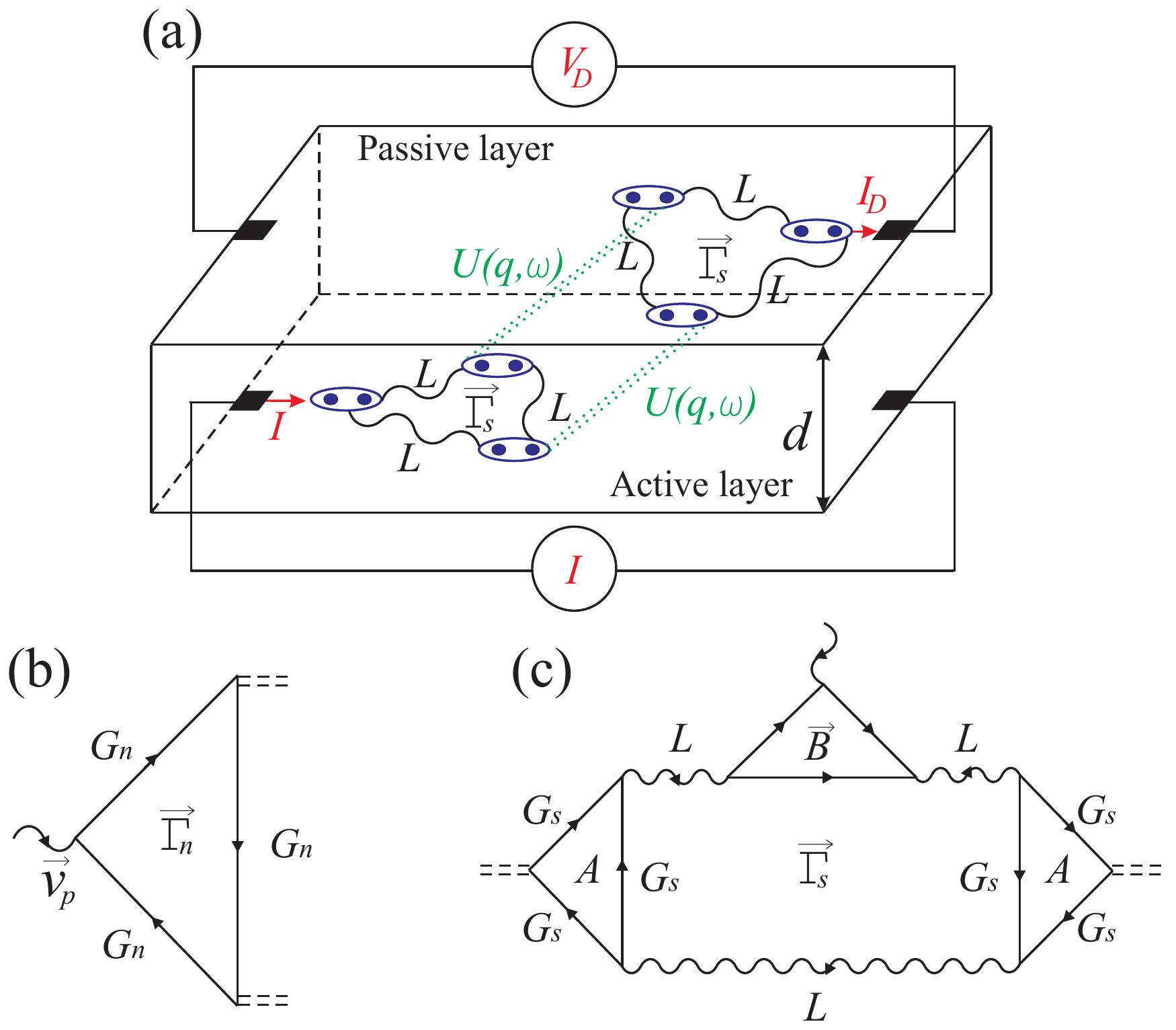}\\
  \caption{(Color online) (a) Schematic representation of the rectification effect
  due to preformed Cooper pairs in a double-layer superconductor Giaever device.
  Diagrams (b) and (c) depict the rectification coefficients $\vec{\Gamma}_{n(s)}$
  in the normal metal and the superconductor, respectively.}\label{fig}
\end{figure}

We assume a screened Coulomb potential as the leading source
of coupling between the layers
\begin{equation}\label{U}
U(q,\omega)=\frac{\pi e^2q}{\kappa_{n(s)}\kappa_s\sinh(qd)}.
\end{equation}
Here we have neglected retardation effects.  $d$ is the spacing between the
layers and $\kappa_{n(s)}=4\pi e^2N_{n(s)}$ is the Thomas-Fermi
screening momentum in the normal metal (superconductor).
Applying a linear response analysis~\cite{Drag-2DEG-Th} to a Giaever
device (Fig.~\ref{fig}a), we have for the drag conductivity
\begin{equation}\label{drag-definition}
\sigma_D\!=\frac{1}{16\pi
T}\!\sum_q\!\int^{+\infty}_{-\infty}\!\!
\frac{d\omega|U(q,\omega)|^2}{\sinh^2\frac{\omega}{2T}}
\vec{\Gamma}_{n(s)}(q,\omega)\vec{\Gamma}_{s}(q,\omega)
\end{equation}
where $\vec{\Gamma}_{n(s)}$ is the nonlinear susceptibility
(rectification coefficient) within each layer. Knowing $\sigma_D$
and inverting the conductivity tensor, one finds the drag resistivity
$\rho_D=\sigma_D/(\sigma_n\sigma_s)$, and finally the drag
resistance $R_D=\rho_D(\ell/w)$, with $\ell$ and $w$ being the length
and width of the two-dimensional strip. In the normal metal, the
rectification coefficient is given by (Fig.~\ref{fig}b)
\begin{eqnarray}\label{Gamma-n}
\vec{\Gamma}_{n}(q,i\omega_1,i\omega_2)=eT\!\!\!\sum_{p,\varepsilon_l,\pm}\vec{v}_p
G_n(p_\pm,i\varepsilon_l)\nonumber\\G_n(p_\mp,i\varepsilon_l\pm
i\omega_1)G_n(p_\mp,i\varepsilon_l\pm i\omega_2)
\end{eqnarray}
with $p_\pm=p\pm q/2$, and $\varepsilon_l=2\pi T(l+1/2)$ being the
fermionic Matsubara frequency. The susceptibility
$\vec{\Gamma}_n(q,\omega)$ which enters
Eq.~\eqref{drag-definition} is obtained from Eq.~\eqref{Gamma-n}
upon appropriate analytic continuation $i\omega_{1,2}\to\omega\pm
i0$ such that
$\vec{\Gamma}_{n}(q,\omega)\equiv\vec{\Gamma}^{RA}_n(q,\omega,\omega)$,
where indices $R(A)$ stand for the retarded (advanced) components in
the complex frequency plane. In the superconducting layer we have
(Fig.~\ref{fig}c)
\begin{eqnarray}\label{Gamma-s}
&&\hskip-.3cm
\vec{\Gamma}_{s}(q,i\omega_1,i\omega_2)=4eT\!\!\!\sum_{Q,\Omega_m,\pm}
\vec{B}(Q_{\pm},i\omega_{1,2},i\Omega_m)\nonumber\\
&&\hskip-.3cm
A(Q,q,i\Omega_m+i\omega_1,i\Omega_m)
A(Q,q,i\Omega_m,i\Omega_m+i\omega_2)\nonumber\\
&&\hskip-.3cm L(Q_\mp,i\Omega_m)L(Q_\pm,i\Omega_m\pm
i\omega_1)L(Q_\pm,i\Omega_m\pm i\omega_2)
\end{eqnarray}
with $Q_\pm=Q\pm q/2$ and $\Omega_m=2\pi m T$ being the bosonic
Matsubara frequency. The scalar (Coulomb) vertex function defining
$\vec{\Gamma}_s$ has the explicit structure
\begin{eqnarray}\label{A}
&&\hskip-.4cm
A(Q,q,i\Omega_m+i\omega_1,i\Omega_m)=
T\sum_{p,\varepsilon_l}G_s(p,i\varepsilon_l)\nonumber\\
&&\hskip-.4cm
G_s(Q-p,i\Omega_m+i\omega_1-i\varepsilon_l)
G_s(p-q,i\varepsilon_l-i\omega_1)
\end{eqnarray}
while the current vertex reads
\begin{eqnarray}\label{B}
&&\hskip-.3cm\vec{B}(Q,i\omega_{1,2},i\Omega_m)=T\!\sum_{p,\varepsilon_l}
\vec{v}_p G_s(p,i\varepsilon_l)\nonumber\\
&&\hskip-.3cm
G_s(p,i\varepsilon_l-i\omega_1+i\omega_2)G_s(Q-p,i\Omega_m+i\omega_1-i\varepsilon_l)
\end{eqnarray}
Similar to the previous case, $\vec{\Gamma}_s(q,\omega)$ entering
Eq.~\eqref{drag-definition} is obtained from Eq.~\eqref{Gamma-s}
after analytic continuation to the real energy axis
$i\omega_{1,2}\to\omega\pm i0$. From Eq.~\eqref{Gamma-s} one sees an
analogy between rectification and the Nernst effect. Indeed, the
Nernst is given by a Feynman graph of three currents connected by
fluctuation propagators similar to Fig.~\ref{fig}c. The rectification
coefficient $\vec{\Gamma}_s$ has the same structure with the only
difference being that two of the current vertices are replaced by a scalar field which
couples to the electromagnetic potential in the other layer. This point
is illustrated in Fig.~\ref{fig}a, which schematically
represents the rectification effect due to fluctuating pairs.
In order to find $\sigma_D$, we need an explicit form for
$\vec{\Gamma}_{n(s)}(q,\omega)$ which we derive below.

\textit{Results}-- We start from Eq.~\eqref{Gamma-n} and convert the
Matsubara sum over the fermionic frequency into the contour integral
$T\sum_{\varepsilon_l}\to\oint_C\frac{d\varepsilon}{4\pi
i}\tanh\frac{\varepsilon}{2T}G_n(\varepsilon)G_n(\varepsilon\pm
i\omega_1)G_n(\varepsilon\pm i\omega_2)$, where the momentum variables
of the Greens functions have been suppressed for brevity. The contour is
a circle with three branch cuts at
$\mathrm{Im}(\varepsilon)=\{0,-\omega_1,-\omega_2\}$.
The outer branches of the
contour that have all three Greens functions of the same causality
do not contribute. The remaining parts, after analytic continuation, give us
in the ballistic case
\begin{equation}
\vec{\Gamma}_n(q,\omega)=\frac{e\omega}{\pi}\sum_{p,\pm}\vec{v}_p\mathrm{Im}[G^R_n(p\mp
q,\varepsilon_F)]|G^R_n(p,\varepsilon_F\pm\omega)|^2
\end{equation}
where we assumed
$\int^{+\infty}_{-\infty}d\varepsilon
\big[\tanh\frac{\varepsilon+\omega}{2T}-\tanh\frac{\varepsilon}{2T}\big]=2\omega$
(low temperature limit).
We next expand the Greens function to linear order in $\vec{q}$ and
perform the remaining momentum integration by approximating
$\sum_p\vec{v}_p(\vec{v}_p\cdot\vec{q})\ldots=
N_n\vec{q}\int^{+\infty}_{-\infty}d\xi_p\frac{\varepsilon_F+\xi_p}{m^*}\ldots$
which accounts for the curvature of the electronic dispersion. Since
$\mathrm{max}\{\omega,q^2/2m^*\}\ll\varepsilon_F$ we find to
leading order
\begin{equation}\label{Gamma-n-fin}
\vec{\Gamma}_n(q,\omega)=4e\frac{D_n\vec{q}}{\varepsilon_F}\frac{N_n\omega}{v_Fq}
\end{equation}
which agrees with earlier calculations~\cite{Drag-2DEG-Th}. One
important remark here is that for the linearized dispersion,
$\xi_p=v_Fp$, rectification vanishes due to electron-hole symmetry.

We proceed with the derivation of $\vec{\Gamma}_s(q,\omega)$,
taking into account a temperature
dependent scattering rate $\gamma_s\simeq T$ appropriate for cuprates
[Eq.~\eqref{G}]. Starting from Eq.~\eqref{Gamma-s} we convert the
bosonic Matsubara sum into the contour integral
$T\sum_{\Omega_m}\to\frac{1}{4\pi i}\oint_c
d\Omega\coth\frac{\Omega}{2T}
A(\Omega+i\omega_1,\Omega)A(\Omega,\Omega+i\omega_2)
L(\Omega)L(\Omega\pm i\omega_1)L(\Omega\pm i\omega_2)$ with three
branch cuts at $\mathrm{Im}(\Omega)=\{0,\mp\omega_1,\mp\omega_2\}$.
After analytic continuation and expansion of the
fluctuational propagator [Eq.~\eqref{L}] to leading
non-vanishing order in the electron-hole asymmetry $\Upsilon_\Omega$, we
find
\begin{eqnarray}\label{Gamma-s-1}
\vec{\Gamma}_s(q,\omega)=\frac{4eN_s}{\pi}|A|^2\vec{B}(q)
\sum_{Q}\int^{+\infty}_{-\infty}d\Omega\,
\Upsilon_\Omega\mathcal{F}(\Omega,\omega)
\nonumber\\
|L^R(Q,\Omega)|^2\mathrm{Re}[L^R(Q,\Omega)]\mathrm{Im}[L^R(Q,\Omega+\omega)]
\end{eqnarray}
where
$\mathcal{F}(\Omega,\omega)=\coth\frac{\Omega+\omega}{2T}-\coth\frac{\Omega}{2T}$.
When deriving this expression, we made use of the following key
observations. (i) Near $T_c$ the characteristic bosonic energies are
smaller than the fermionic ones $\{DQ^2,\Omega,\omega\}\sim T-T_c\ll T$.
This allows us to compute the vertex functions [Eqs.~\eqref{A}-\eqref{B}]
by neglecting their dependence on the bosonic modes. This step is
legitimate since both $A$ and $\vec{B}$ contain a fermionic loop with
$\varepsilon\sim T$. (ii) Under the above approximation, one then
finds from Eqs.~\eqref{A} and $\eqref{B}$ with the help of
Eq.~\eqref{G}
\begin{eqnarray}
|A|^2=\frac{N^2_s}{\Delta^2}\alpha^2(\gamma_s/\Delta),\quad
\vec{B}(q)=-\frac{\pi N_sD_s\vec{q}}{4\Delta}\beta(\gamma_s/\Delta)
\end{eqnarray}
where scaling functions
\begin{eqnarray}
&&\hskip-.5cm \alpha(z)=\frac{1}{16\pi}\left[\frac{1}{\sqrt{1+z^2}}
+3\ln\left(\frac{1+\sqrt{1+z^2}}{|z|}\right)\right]\\
&&\hskip-.5cm \beta(z)=\frac{\pi^2\xi^2}{24\eta}
\left[\frac{z^2+2}{z\sqrt{z^2+1}}-1\right]\label{beta}
\end{eqnarray}
account for the suppression of the vertex due to the pseudogap~\cite{LNV}.
In Eq.~\eqref{beta}, $\xi=\hbar v_F/\pi\Delta$ and $\eta=\pi
D/8\Delta$. (iii) We also exploited the fact that $Q\gg q$ and
set $q\to0$ in the propagators since rectification is determined by
small momentum transfer between the layers, namely $Dq^2\sim
(T-T_c)/(\kappa_{n(s)} d)\ll T-T_c\sim DQ^2$. (iv) Our final comment
regarding Eq.~\eqref{Gamma-s-1} concerns the electron-hole asymmetry. In
general $\Upsilon_\Omega$ enters both the fluctuational propagator $L$
and the scalar vertex $A$ which, in fact, is a consequence of gauge
invariance~\cite{Drag-2DEG-Th,AHL}. We find that the most singular
in $T-T_c$ contribution to $\sigma_D$ comes from keeping
$\Upsilon_\Omega$ in $L$ while taking the scalar vertex as the bare one.
The technical reasoning is that by expanding $L$ in
$\Upsilon_\Omega$, one effectively raises the power of the fluctuational
propagator that in the end translates into an extra factor of
$\epsilon$ in $\sigma_D$ after energy integrations. Conversely,
keeping $\Upsilon_\Omega$ in $A$ would also lead to a nonvanishing
contribution to $\vec{\Gamma}_s$, but with a subleading dependence on
$T-T_c$ near the transition.

We can now simplify the integrand in Eq.~\eqref{Gamma-s-1} even
further by exploiting the  separation of the energy scales
$\{\Omega,\omega\}\ll T$, thus approximating
$\mathcal{F}(\Omega,\omega)\approx-\frac{2T\omega}{\Omega(\Omega+\omega)}$
and using Eq.~\eqref{L} which gives us
\begin{eqnarray}
&&\hskip-.5cm \vec{\Gamma}_s(q,\omega)= -\frac{32eT^2}{\pi^2
N^3_s}\vec{B}(q)|A|^2\frac{\partial\ln
\Delta}{\partial\varepsilon_F}\nonumber\\
&&\hskip-.5cm \sum_Q\int^{+\infty}_{-\infty}d\Xi_\Omega
\frac{\Lambda_Q\Xi_\omega}{[\Lambda^2_Q+\Xi^2_\Omega]^2
[\Lambda^2_Q+(\Xi_\Omega+\Xi_\omega)^2]}
\end{eqnarray}
Here we introduced dimensionless variables $\Xi_\Omega=\pi\Omega/8T$
and $\Lambda_Q=\pi D_sQ^2/8T+\epsilon$. The remaining energy and
momentum integrations can be completed analytically with the final
result
\begin{eqnarray}\label{Gamma-s-fin}
\vec{\Gamma}_s(q,\omega)=-\frac{4eT^2}{\pi^2D_sN^3_s\epsilon^3}
\vec{B}(q)|A|^2\frac{\partial\ln\Delta}{\partial\varepsilon_F}
\frac{\omega}{4+(\omega\tau_{GL})^2}
\end{eqnarray}
where the Ginzburg-Landau time $\tau_{GL}=\pi/(8T\epsilon)$ was
introduced. By combining Eqs.~\eqref{Gamma-n-fin} and
\eqref{Gamma-s-fin} in Eq.~\eqref{drag-definition}, we find the drag
conductivity due to fluctuational rectification by preformed Cooper
pairs in the non-symmetric normal metal-superconductor transformer
(restoring $\hbar$)
\begin{equation}\label{sigma-ns}
\frac{\sigma_D}{\sigma_Q}=\frac{\pi^2r^2_s}{30}
\frac{\partial\Delta}{\partial\varepsilon_F}
\frac{\lambda_{ns}(d)f(T)}{\ln^2\frac{T}{T_c}}
\end{equation}
where $\sigma_Q=\frac{e^2}{\hbar}$,
$\lambda_{ns}(d)=\frac{(D_n/v_Fd)}{(\kappa_nd)^2(\kappa_sd)^2}$,
$f(T)=\left(\frac{T}{\Delta}\right)^4\alpha^2\beta$  and
$r_s=\frac{e^2}{\hbar v_F}$. In the immediate vicinity of the
transition, $T-T_c\ll T_c$, where $f(T)$ is constant, the rectification
has a strong power-law enhancement,
$\sigma_D\propto\left(\frac{T_c}{T-T_c}\right)^{2}$, which
physically corresponds to an Aslamazov-Larkin (AL) paraconductivity
effect~\cite{AL}. Notice that $\sigma_D$ is more singular than the AL
conductivity $\sigma_{AL}\propto\frac{T_c}{T-T_c}$. This situation
is similar to the Hall effect in superconductors \cite{book,AHL} since
both the drag and Hall conductivities are sensitive to the electron-hole
asymmetry and require an expansion of $L$ in $\Upsilon_\Omega$, while the AL
conductivity is not. Further away from the transition,
the vertex suppression due to the pseudogap, where $\alpha\propto\Delta/T$,
$\beta\propto(\Delta/T)^2$ and $f(T)\sim1$, results in a moderate
decay of $\sigma_D\propto\frac{1}{\ln^2\frac{T}{T_c}}$. Screening
effects on the Coulomb potential provide an additional suppression
of $\sigma_D$ for large separation between the layers.

In the case when both layers are superconductors, we use two vertices
from Eq.~\eqref{Gamma-s-fin} in Eq.~\eqref{drag-definition} and find
\begin{equation}\label{sigma-ss}
\frac{\sigma_D}{\sigma_Q}=\frac{15\zeta(5)r^2_s}{32}\frac{\partial\Delta_1}{\partial\varepsilon_F}
\frac{\partial\Delta_2}{\partial\varepsilon_F}
\frac{\lambda_{ss}(d)g(T)}
{\ln^2\frac{T}{T_{c1}}\ln^2\frac{T}{T_{c2}}
\left[\ln\frac{T}{T_{c1}}+\ln\frac{T}{T_{c2}}\right]}
\end{equation}
where
$\lambda_{ss}(d)=\frac{(\xi_1\xi_2/d^2)}{(\kappa_{s1}d)^2(\kappa_{s2}d)^2}$,
$g(T)=\left(\frac{T^2}{\Delta_1\Delta_2}\right)^3\alpha^2_1\alpha^2_2\beta_1\beta_2$
with shorthand notation
$\{\alpha,\beta\}_{1,2}=\{\alpha,\beta\}(\gamma_s/\Delta_{1,2})$.
This result shows a similar but potentially more singular behavior near the transition as
Eq.~\eqref{sigma-ns}, however, with a much faster power-law decay at higher
temperatures, $\sigma_D\propto\frac{1}{T^2\ln^5T}$, since
$g(T)\propto(\Delta_1\Delta_2/T^2)$ for
$T>\mathrm{max}\{T_{c1},T_{c2}\}$. This high-$T$ suppression
mechanism due to the pseudogap was recently proposed as an explanation of the fast
fall-off of the Nernst signal in underdoped cuprates~\cite{LNV}.

\textit{Summary}-- In conclusion, we developed a
theory of a Giaever transformer based on bi-layers of
cuprate superconductors, assuming that the leading
source of rectification originates from fluctuating Cooper
pairs. The coupling between the layers is provided by a screened
Coulomb interaction. The drag signal shows a strong singularity near the
superconducting transition due to the paraconductivity effect, while the
pseudogap leads to a faster drop-off at higher temperatures.
Such a characteristic temperature dependence may distinguish drag due
to preformed pairs from that due to vortices. In particular,
Ref.~\onlinecite{Shimshoni} predicted that the drag resistivity $\rho_D$ in
a normal-superconductor bilayer due to the inductive coupling effect
of vortices should follow the flux flow conductivity $\sigma_v$ of the
superconductor, $\rho_D\propto(\hbar/e^2)^2\sigma_v$. Clearly, this
result is very different from the contribution to $\rho_D$
originating from fluctuating Cooper pairs [Eq.~\eqref{sigma-ns}]. The general
condition to have a vortex drag effect is to have an inhomogeneous
magnetic field in the barrier region (otherwise, the vortices
between the two layers cannot `find' each other). This means that
the vortices must be well defined. As a consequence, any vortex drag
regime above $T_c$ should have a narrow temperature range. This is
especially true for cuprates, given the weak coupling of vortices
between layers (i.e., pancake vortices). That coupling gets even
weaker for underdoped compounds, which is exactly the regime we are
interested in.

This work was supported by the US DOE, Office of Science, under
contract DE-AC02-06CH11357 and by the Center for Emergent
Superconductivity, an Energy Frontier Research Center funded by the
US DOE, Office of Science, under Award No.~DE-AC0298CH1088.

\end{document}